Correspondence to:
J. P. Reistad,
jone.reistad@uib.no

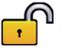

# Separation and Quantification of Ionospheric Convection Sources: 2. The Dipole Tilt Angle Influence on Reverse Convection Cells During Northward IMF

J. P. Reistad[1], K. M. Laundal[1], N. Østgaard[1], A. Ohma[1], E. G. Thomas[2], S. Haaland[1,3], K. Oksavik[1,4], and S. E. Milan[1,5]

[1]Birkeland Centre for Space Science, University of Bergen, Bergen, Norway, [2]Thayer School of Engineering, Dartmouth College, Hanover, NH, USA, [3]Max Planck Institute for Solar System Research, Göttingen, Germany, [4]Arctic Geophysics, University Centre in Svalbard, Longyearbyen, Norway, [5]Department of Physics and Astronomy, University of Leicester, Leicester, UK

**Abstract** This paper investigates the influence of Earth's dipole tilt angle on the reverse convection cells (sometimes referred to as lobe cells) in the Northern Hemisphere ionosphere during northward IMF, which we relate to high-latitude reconnection. Super Dual Auroral Radar Network plasma drift observations in 2010–2016 are used to quantify the ionospheric convection. A novel technique based on Spherical Elementary Convection Systems (SECS) that was presented in our companion paper (Reistad et al., 2019, https://doi.org/10.1029/2019JA026634) is used to isolate and quantify the reverse convection cells. We find that the dipole tilt angle has a linear influence on the reverse cell potential. In the Northern Hemisphere the reverse cell potential is typically two times higher in summer than in winter. This change is interpreted as the change in interplanetary magnetic field-lobe reconnection rate due to the orientation of the dipole tilt. Hence, the dipole tilt influence on reverse ionospheric convection can be a significant modification of the more known influence from $v_{sw}B_z$. These results could be adopted by the scientific community as key input parameters for lobe reconnection coupling functions.

## 1. Introduction

Magnetic reconnection allows energy from the solar wind and its embedded interplanetary magnetic field (IMF) to enter and be distributed within the magnetosphere system. In terms of energy transport in the system, the Dungey cycle (Dungey, 1961; opening of magnetic flux on the dayside and subsequent closure on the nightside) is of most importance. For this cycle, a strong degree of symmetry between the two polar regions is required by the Maxwell equation $\nabla \cdot \vec{B} = 0$, requiring both polar caps to contain the same amount of open flux.

However, reconnection can take place at other locations in the magnetosphere system, leading to plasma circulations that are not necessarily restricted by the north-south symmetry constraints for the Dungey cycle (Wilder et al., 2013). This is the case during lobe reconnection, where the IMF reconnects with open field lines at the high-latitude magnetopause just tailward of the cusp, as schematically illustrated in Figure 1a. The lobe reconnection process is entirely independent in the two hemispheres. However, it is possible that an IMF field line can simultaneously reconnect with the lobe in both hemispheres leading to closure of open flux (Imber et al., 2006), known as dual lobe reconnection. Based on analysis of global magnetospheric magnetohydrodynamic modeling, Watanabe et al. (2005) and (Watanabe & Sofko, 2009a, 2009b) have pointed out additional possible reconnection geometries during northward IMF, all occurring on high latitudes, tailward of the cusps. Although the relative importance of these other reconnection geometries are still unknown, we will discuss our results in light of this framework in section 4. Throughout the text we will use the term "high-latitude reconnection" when discussing any of these possible reconnection processes associated with northward IMF. When using the term "lobe reconnection," we specifically refer to the IMF-lobe reconnection process as illustrated in Figure 1a.

Although the energy transport related to lobe reconnection is usually much less than the one associated with the Dungey type reconnection, the region of influence is also much smaller, usually limited to above 80° MLAT on the dayside (06–18 MLT) in the ionosphere. However, the redistribution of open flux during lobe reconnection has also been shown to introduce asymmetries on closed field lines (Tenfjord et al., 2018). Within the polar cap, strong disturbances due to the lobe reconnection process are observed (e.g., Burch et al., 1980; Friis-Christensen & Wilhjelm, 1975; Wilder et al., 2010). Furthermore, as the plasma circulation within the polar cap due to lobe reconnection is not bound by the same north-south symmetry constraints as the flows initiated by the Dungey cycle, its dependence on IMF and solar wind parameters might be different in the two hemispheres. One obvious geometric difference is the different inclination toward the Sun of the two polar regions, as quantified by the dipole tilt angle; see Figure 1a. The local conditions determining the reconnection rate are the shear angle between the two magnetic domains and the rate of flux transport toward the reconnection line. Since the dipole tilt angle can change by almost 70° between its extreme values around the solstices, this is a likely source of variability of the reconnection rate. The purpose of this paper is to give an estimate of the influence on the lobe reconnection rate due to variations of the dipole tilt angle.

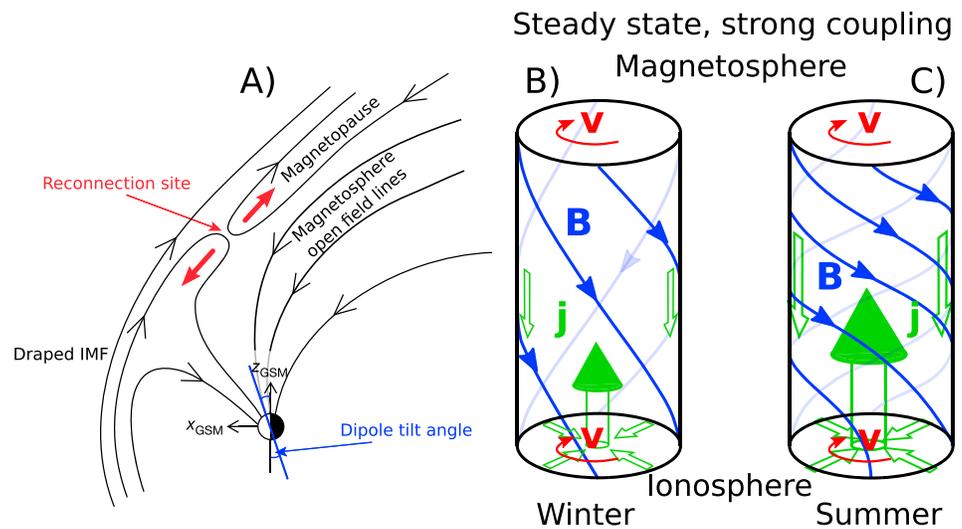

**Figure 1.** (a) A sketch of the lobe reconnection geometry at the Northern Hemisphere high-latitude magnetopause when the Earth's dipole (blue axis) is inclined toward the Sun (northern summer, positive tilt). (b and c) A conceptual view of the magnetosphere-ionosphere coupling in a steady state where the ionosphere is considered to respond only to the magnetospheric forcing. The ionosphere is assumed to have a homogeneous conductivity, being low in panel (b) representing a dark ionosphere and higher in panel (c) representing a sunlit ionosphere. The same level of magnetospheric circulation is imposed in both panels (b) and (c), resulting in equal plasma circulation in the ionosphere as illustrated with the red arrows. The higher conductivity levels in (c) will lead to a stronger shear in $\vec{B}$ (blue lines) as a result of increased ionospheric friction, leading to a larger $j_\parallel$ (large green arrow) in the sunlit ionosphere. Panels (b) and (c) are reproduced from Paschmann et al. (2002) Figure 3.7. IMF = interplanetary magnetic field.

There are good physical reasons to relate observations of sunward ionospheric convection in the dayside polar cap during northward IMF to lobe reconnection. The observation of strong sunward convection at high latitudes was in fact the strongest argument for the existence of lobe reconnection. Although the existence of lobe reconnection was proposed by Dungey (1963), direct observations at the magnetopause were not presented until decades later (Gosling et al., 1991; Kessel et al., 1996). The reason why signatures of sunward ionospheric convection in response to lobe reconnection was expected is due to the strong coupling between the magnetosphere and polar cap ionosphere. In a steady state description, one can directly relate the driver, for example, lobe reconnection initiating circulation of plasma in the lobes, to the ionospheric response, namely, a corresponding circulation of plasma within the dayside polar cap. This situation is schematically illustrated in Figures 1b and 1c for two different ionospheric conditions: (b) local winter where the ionosphere has low and uniform conductivity and (c) local summer where the ionosphere has high and uniform conductivity. Figures 1b and 1c are adopted from Paschmann et al. (2002) Figure 3.7. Note that the lobe reconnection rate is assumed to be the same in the two cases, hence the identical red arrows of plasma circulation on the magnetospheric side. As the stresses from the twisting of the magnetic field at the magnetospheric side will propagate toward the ionosphere, the F region ionospheric plasma will start to rotate in response, leading to a $\nabla \cdot \vec{E}$ within the rotating tube since $\vec{E} = -\vec{v} \times \vec{B}$ in the F region ionosphere and above in this strong coupling scenario. In the steady state, strong coupling case (what is shown in Figures 1b and 1c), the ionospheric convection will be at a rate corresponding to a constant shear in $\vec{B}$, hence independent of ionospheric conductivity. The amount of shear in $\vec{B}$, proportional to $j_\parallel$ (green arrow), is related to the force needed to move the ionospheric footprints. During local winter (Figure 1b), the ionospheric conductivity and hence the frictional force are low, leading to a small shear in $\vec{B}$. During local summer conditions (Figure 1c), the conductivity is higher. The increased friction lead to a larger shear in $\vec{B}$ and hence $j_\parallel$ to accommodate the steady state during the same levels of magnetospheric circulation. In this discussion we have assumed that a decoupling between the magnetosphere and ionosphere due to $E_\parallel$ is negligible inside the dayside polar cap. According to statistical studies of particle precipitation, $E_\parallel$ associated precipitation is on average mainly confined to the duskside oval (Newell et al., 1996, 2009), enabling a ground-based investigation of the origin of plasma convection in the magnetosphere. Despite being limited by assumptions about the coupling between the polar cap ionosphere and the magnetosphere, the ground-based nature of this study benefits from being able to observe the footprint of the entire magnetospheric region where these interactions take place.

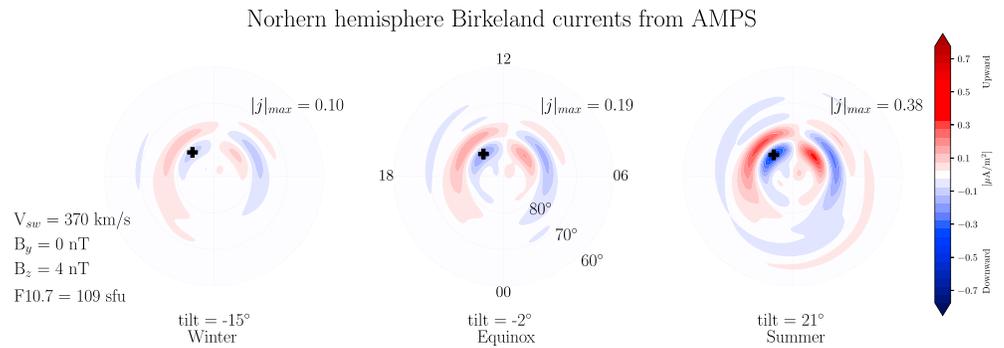

**Figure 2.** Average Magnetic field and Polar current System (AMPS) model values of Birkeland currents during purely northward interplanetary magnetic field for three different values of the dipole tilt angle corresponding to northern winter, equinox, and summer conditions. Model parameters are chosen to reflect the average conditions in Figure 3.

Birkeland currents mapping to the magnetospheric lobe circulation region cannot be directly used to infer the strength of the lobe reconnection rate for different orientations of the dipole tilt angle, as the currents are highly influenced by the ionospheric conductivity, which in turn depend strongly on the dipole tilt angle. An example of this, which will be used as reference for our results, is shown in Figure 2. Here, Birkeland currents from the Average Magnetic field and Polar current System (AMPS) model (Laundal, Finlay, et al., 2018) are shown. AMPS is an empirical model of the perturbation magnetic field derived from low Earth orbiting satellites, from which the full 3-D ionospheric current system can be calculated. The model is parameterized by the external parameters $v_{sw}$, IMF $B_y$, IMF $B_z$, dipole tilt angle, and F10.7 index and is designed to reflect the influence of the solar wind-magnetosphere interactions taking place on the dayside. Figure 2 shows the Birkeland currents from the Northern Hemisphere during purely northward IMF for three different dipole tilt values corresponding to local winter, equinox, and summer conditions. The model parameter input is printed on the figure and is chosen to reflect the average conditions of the statistics presented in Figure 3. The NBZ (northward $B_z$; Iijima, 1984) current system is clearly seen, located poleward of the dayside region 1 currents, at 80–85° MLAT. These currents we attribute to high-latitude reconnection, and the maximum absolute value within 80° MLAT is printed in each panel, corresponding to the location of the "+" symbol. Using this value as a proxy of the intensity of the current system associated with high-latitude reconnection, it can be seen, as also qualitatively shown earlier (e.g., Green et al., 2009; Laundal, Finlay, et al., 2018; Weimer, 2001), that the NBZ currents increase by a factor of ~4 going from tilt = −15° to tilt = 21°. We will later return to this average influence of dipole tilt on the NBZ currents, as our results will put constraints on how much of this variation we attribute to an increase in high-latitude reconnection and how much is related to the increased solar-induced conductivity.

Earlier studies have found strong evidence that lobe reconnection is more efficient in the hemisphere inclined toward the Sun, that is, the local summer hemisphere (Crooker & Rich, 1993; Frey et al., 2004; Koustov et al., 2017; Østgaard et al., 2018; Wilder et al., 2010; Yakymenko et al., 2018), from observing the sunward ionospheric convection velocity in the dayside polar cap during northward IMF. However, most studies have focused on the dependence on the solar wind electric field, $E_{sw} = v_{sw} B_T$, where $B_T$ is the trans-verse component of the IMF, $B_T = \sqrt{B_y^2 + B_z^2}$ (Koustov et al., 2017; Sundberg et al., 2009; Wilder et al., 2009; 2010; Yakymenko et al., 2018), likely to be the most important controlling factor on the lobe reconnection rate during purely northward IMF. These studies all suggest a linear dependence of the ionospheric sunward convection speed in the dayside polar cap to $E_{sw}$ during northward IMF; however, a saturation effect is observed when $E_{sw} \gtrsim 3$ mV/m.

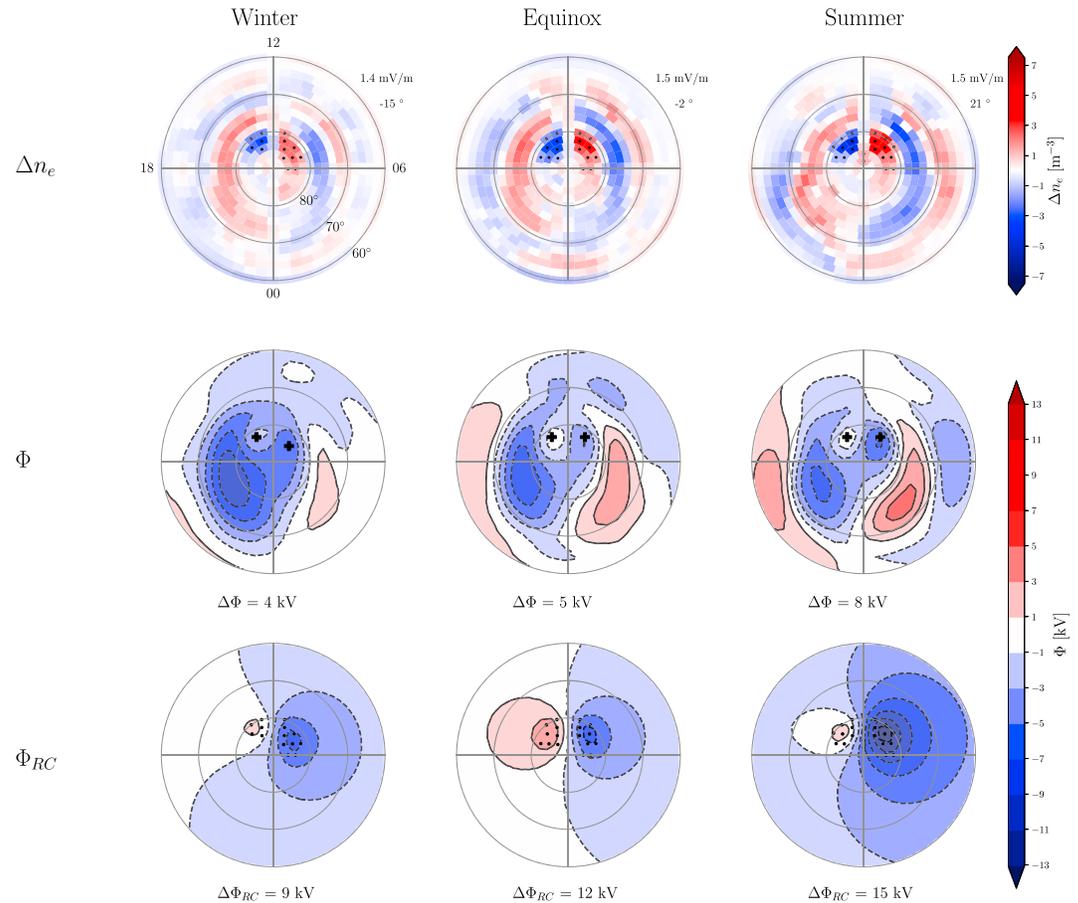

**Figure 3.** Inferred reverse convection potential $\Phi_{RC}$ from the Northern Hemisphere during northward interplanetary magnetic field (IMF) for three different dipole tilt intervals based on Super Dual Auroral Radar Network line-of-sight measurements: (left column) winter (dipole tilt less than $-10°$), (middle column) equinox (dipole tilt within $\pm 10°$), and (right column) summer (dipole tilt $>10°$). The upper row shows the inferred $\Delta n_e$ from the Spherical Elementary Convection Systems analysis, with reverse convection grid cells indicated with black dots. The middle row shows the electric potential $\Phi$. The bottom row shows the electric potential resulting from the selected reverse convection grid cells only. This potential is what we relate to the lobe reconnection rate and is printed below each panel.

A key difference of the present study compared to other studies on the seasonal differences of ionospheric convection during northward IMF is the focus on the magnetic flux transport rate in the ionosphere, which we assume is directly related to the high-latitude reconnection rate due to the strong coupling between the two regions. Previous studies have mainly focused on the ionospheric sunward convection speed. However, the study by Chisham et al. (2004) is an exception, where the lobe reconnection rate is inferred from ionospheric observations in a case study, also assuming the strong coupling between the ionosphere and the reconnection region. In order to arrive at a lobe reconnection coupling function, which is currently not existing in literature to our knowledge, the ionospheric convection needs to be translated into a magnetic flux transport rate; that is, the magnetic field strength and the distance where this convection exists must be taken into account. The present study aims to quantify the contribution from the dipole tilt angle to such a coupling function. Our findings also suggest that the high-latitude reconnection rate depends linearly on the dipole tilt angle and is typically two times higher during summer versus winter for purely northward IMF.

The present paper takes advantage of a novel technique described in our companion paper *Separation and quantification of ionospheric convection sources: 1. A new technique*, referred to as Paper I. In Paper I we show how this technique can be used to separate the different sources of the ionospheric convection, enabling us in particular to estimate the magnetic flux transport rate (the potential) associated with the high-latitude reconnection processes. In the following section we describe the underlying convection data set that the method in Paper I is applied to. Section 3 presents the results during northward IMF for different orientations of the Earth's dipole axis. Sections 4 and 5 discuss and summarize the results.

## 2. Method

We here describe the underlying data used to make the convection maps, the selection of external driving conditions, and finally a brief summary of the main steps of the Spherical Elementary Convection Systems (SECS) technique described in Paper I.

### 2.1. Ionospheric Convection Data Set

The ionospheric convection data set used in this study is from the Super Dual Aurora Radar Network (SuperDARN; Chisham et al., 2007; Greenwald et al., 1995). The LOS (line-of-sight) ionospheric plasma velocity is deduced from the Doppler shift of the backscattered echo, caused by decameter-scale magnetic field-aligned irregularities in the electron density. Following the procedure of Thomas and Shepherd (2018), 7 years (2010–2016) of LOS velocity data from all available Northern Hemisphere SuperDARN radars are binned onto an equal-area MLAT/MLT grid with spatial resolution of ∼100 km and temporal resolution of 2 min (Ruohoniemi & Baker, 1998). Data from ranges less than 800 km are excluded to prevent contamination by lower-velocity $E$ region echoes. We have also removed velocity data obtained from ranges further away than 2,000 km to reduce the likelihood of geolocation inaccuracies associated with multihop HF radio propagation. Finally, measurements collected during nonstandard radar operating modes are discarded. By doing this, we ensure that only the highest-quality radar data are considered and also allowing for close comparison and validation with the statistical results presented by Thomas and Shepherd (2018), as done in Paper I.

### 2.2. Data Selection Based on Driving Conditions

Since we are interested in describing the ionospheric convection during northward IMF, it is important that we select data during periods when the IMF has been stable and northward for some time, to avoid contamination from flows initiated during southward IMF. We identify intervals of stable IMF using the bias filtering technique (Haaland et al., 2007). With this technique, randomly oriented IMF vectors would lead to bias vector length of 0. For increasingly stable IMF, the length of the bias vector approaches 1. Similar to Haaland et al. (2007), we calculate the bias vector length at a given time based on a 30-min rolling interval including the previous 20 min and the following 10 min using 1-min OMNI data (King & Papitashvili, 2005) and require the length of the bias vector to be >0.96 to be considered stable. This stability criterion is fulfilled 55% of the time in the analysis period. In addition to the IMF stability criterion we average the 1-min OMNI IMF and solar wind observations, representing the conditions at the bow shock, over the previous 20 min. This is to better represent the average conditions that participate in the dayside and lobe reconnection process (Laundal, Finlay, et al., 2018). LOS convection data from SuperDARN are then selected when the IMF clock angle is in the interval [−30°, 30°]. Note that when quantifying the IMF stability, only stability in the YZ plane is considered. We also note that an inherent limitation of the data selection based on upstream solar wind and IMF is the effect of draping of the IMF through the magnetosheath (Sibeck et al., 1990). This will probably account for some of the large spread of the ionospheric convection observations. Only a dedicated in situ study of the lobe reconnection rate would be able to overcome this challenge.

To reduce the known variability of lobe reconnection rate on the solar wind velocity and the magnitude of the northward component of the IMF, we only consider SuperDARN observations when $E_{sw} = v_{sw}B_T$ is between 1 and 2 mV/m, where $B_T$ is the transverse magnitude of IMF. This corresponds to an interval around the peak occurrence of $E_{sw}$ and is satisfied 30% of the time.

The last selection parameter is the dipole tilt angle, defined as the angle between the centered magnetic dipole axis and the GSM Z axis, in the GSM XZ plane. By convention, positive values correspond to northern summer. Since higher-order terms in the multipole expansion of the Earth's magnetic field decrease faster than the dipole term, the dipole tilt angle describes to a large extent the geometric north-south asymmetries imposed from the earthward side of the solar wind-magnetosphere interactions. We have used the dipole coefficients from the International Geomagnetic Reference Field model (Thébault et al., 2015) to determine the dipole orientation that corresponds to the time of SuperDARN observations. This study focuses mainly on three intervals of the dipole tilt angle corresponding to the three main regimes of solar illumination in the Northern Hemisphere: winter (less than $-10°$), equinox (within $\pm10°$), and summer ($>10°$).

### 2.3. Summary of Data Processing

The methodology of the SECS description of the average ionospheric convection is described in Paper I. Here we summarize the main steps of the technique to give a brief background. We describe the convection electric field above 60° MLAT as a sum of the electric field contributions from 480 *nodes* distributed uniformly across the MLT/MLAT domain at fixed locations. Each node has its own curl-free electric field associated with it, pointing away from or toward the node along the spherical surface (ionosphere). In vicinity of the node, the magnitude of the node electric field is approximately proportional to $1/r$, where $r$ is the distance to the node, and is scaled by an amplitude specific for that node. In this specific description, the convection electric field is determined by the 480 node amplitudes, which are estimated by an inversion process based on the observed LOS plasma velocities during the specific selection conditions.

The input data to the SECS analysis are the LOS SuperDARN convection velocities as described in section 2.1 selected during the conditions described in section 2.2. To relate the plasma velocities to an electric field, we also need the value of the magnetic field at each measurement location, found using the International Geomagnetic Reference Field model. To overcome challenges due to an uneven spatial distribution of a very large number of observations, typically $10^6$ LOS observations, we have used an intermediate step in the inversion for the SECS amplitudes. Instead of the direct inversion of a ($\sim 10^6$, 480) size matrix, we compute binned average $E$ fields from the LOS observations on a new grid. This intermediate step is described in detail in section 2.5 in Paper I.

As shown in Paper I, it is the node amplitudes that describe the ionospheric convection in the SECS representation. Paper I presents how to calculate the convection electric field and potential directly from the amplitudes. More importantly, Paper I shows that the values and distribution of the node amplitudes contain information about the magnetospheric sources of the ionospheric convection field, as the values of the node amplitudes reflect the local contributions to the ionospheric convection. Equation 14 in Paper I shows that the node amplitude is proportional to the divergence of the convection electric field. Hence, for a homogeneous convection field, the amplitudes are $\sim 0$, while for a region with structure in the convection, indicating a structure in the magnetospheric convection due to, for example, reconnection, the local node amplitudes increase in absolute value.

In our approach to determine $\vec{E}_{SECS}$, we do not impose any constraints using statistical fill-in data from an empirical model, as is usually the case when instantaneous global convection patterns are derived. This is possible as our analysis is not used to describe a specific event or time interval, but rather the average large scale plasma circulation during specific IMF clock angle, $E_{sw}$, and dipole tilt interval ranges. The large database of SuperDARN data from the years 2010–2016 has no need for fill-in data to calculate the binned average $\vec{E}$ in any grid cell. We use a weakly imposed boundary condition at our low-latitude boundary by including synthetic observations of zero electric field at 59° MLAT in the inversion for $\vec{E}_{SECS}$; see Paper I for details. This is similar to the use of the Heppner-Meynard boundary (Shepherd & Ruohoniemi, 2000) in the SuperDARN map potential technique (Ruohoniemi & Greenwald, 2005).

## 3. Seasonal Variation of the Lobe Cell Circulation

We here infer, using the technique described in Paper I, the reverse convection potential in the Northern Hemisphere during purely northward IMF for different values of the Earth's dipole tilt angle. As mentioned in section 2.2, we also keep $E_{sw} \in [1, 2]$ mV/m, as it is known to be an important controlling parameter for the lobe reconnection rate (Koustov et al., 2017; Sundberg et al., 2009; Wilder et al., 2009). This interval is chosen because of data coverage while still limiting the externally driven variations in the lobe reconnection rate due to $E_{sw}$.

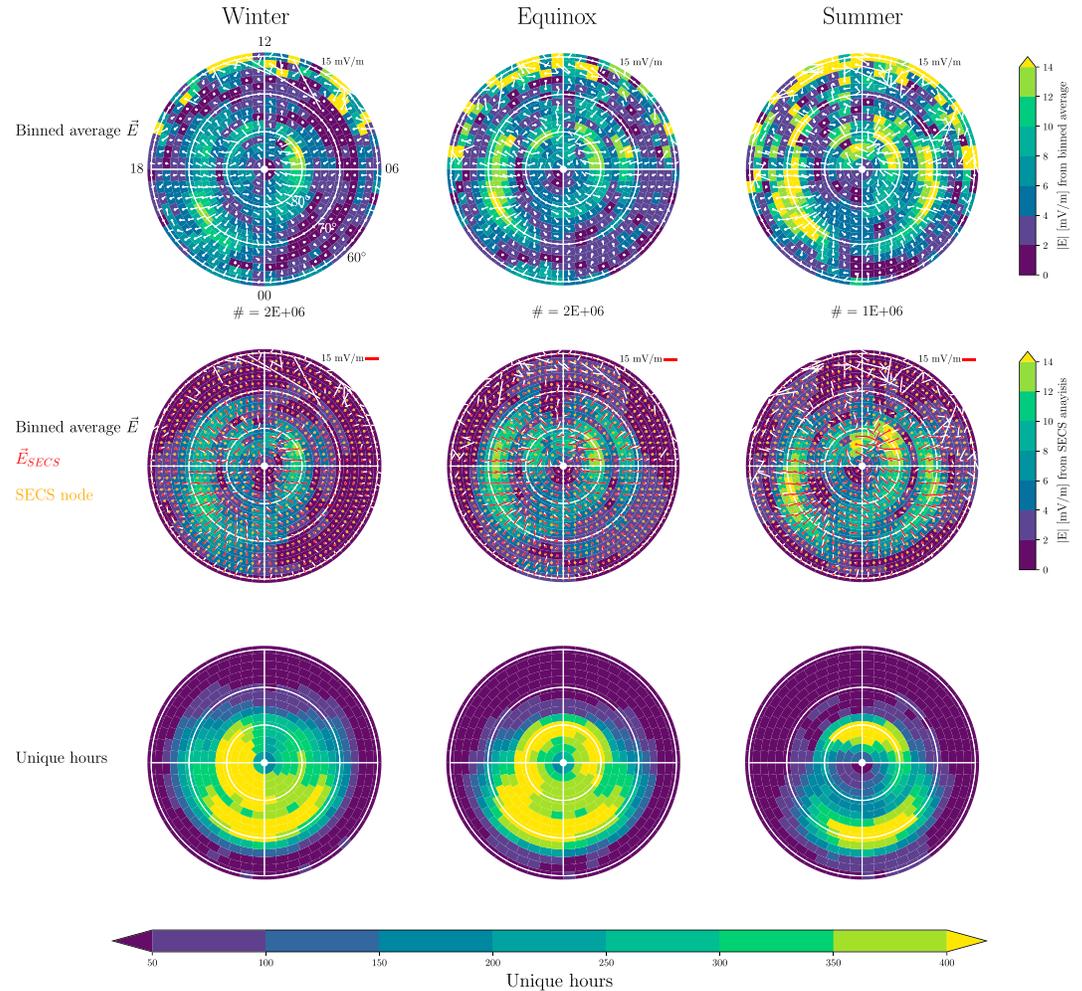

**Figure 4.** Statistics of the Super Dual Auroral Radar Network data used to produce Figure 3. The three columns refer to the same dipole tilt intervals as in Figure 3. The upper row shows the magnitude of the binned average $\vec{E}$ as color and corresponding vectors as white pins from the center of the bin. The middle row is the magnitude of the estimated curl-free $\vec{E}_{SECS}$ and its associated vector shown as red pins. The bottom row is the data coverage shown as number of unique hours of observations in each bin. SECS = Spherical Elementary Convection Systems.

Figure 3 shows the results for three different intervals of the dipole tilt angle. The three columns correspond to the dipole tilt intervals [−35°, −10°] (winter), [−10°, 10°] (equinox), and [10°, 35°] (summer). The top row shows $\Delta n_e$ from the SECS analysis on the SECS node grid. $\Delta n_e$ is the charge density expressed as the number of excess electrons per cubic meter, needed to maintain the electric field, found using Gauss law; see equation 14 in Paper I. We identify the grid cells associated with the reverse convection as done in Paper I, namely, using a threshold value of $|\Delta n_e| > 1$ electrons per cubic meter inside the dayside polar cap (MLAT $\geq 80°$, MLT $\in [6, 18]$). These identified grid cells are highlighted as black dots in the top and bottom rows of Figure 3. The average $E_{sw}$ in the dayside polar cap of the underlying data is shown next to the polar plot at 09 MLT, indicating no significant bias in $E_{sw}$ between the different tilt angle intervals. The corresponding average dipole tilt angle is printed at 08 MLT. We also show the electric potential $\Phi$ in the second row and the potential from the reverse convection nodes only, $\Phi_{RC}$, in the bottom row in Figure 3.

As discussed in Paper I, we argue that the segmentation of the convection, where we express the potential related to the reverse convection cells only (bottom row in Figure 3), is more directly addressing the source of the reverse convection circulation than the potential difference inferred from looking at $\Phi$ in the second row of Figure 3, where the potential difference is calculated from the locations of the maximum and minimum $\Phi$ in the region MLAT $\geq 80°$ and MLT $\in [06, 18]$, indicated with bold "+" symbols in the middle row in Figure 3. However, a significant increase in the inferred reverse convection potential is observed in both cases. Judging from the potentials inferred from the reverse convection nodes only, the reverse convection potential increases from 9 kV for dipole tilt $\in [-35°, -10°]$ to 15 kV for tilt $\in [10°, 35°]$.

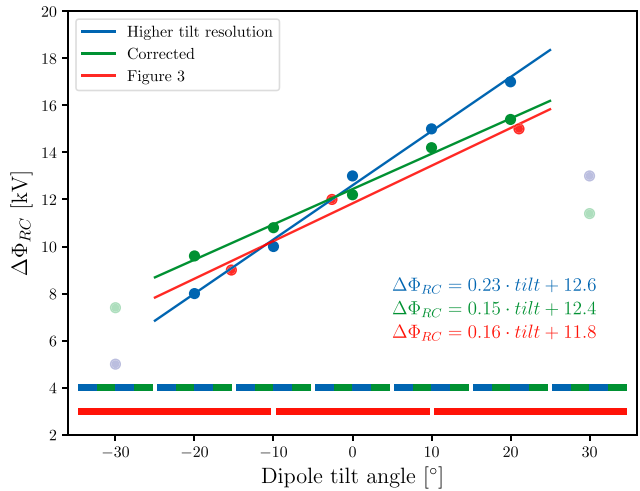

**Figure 5.** A summary of the reverse convection potential difference $\Delta\Phi_{RC}$ versus dipole tilt angle. (Blue) The same analysis as presented in Figure 3 but for smaller dipole tilt intervals, indicated by the horizontal bars at the bottom of the figure (blue/green). $\Delta\Phi_{RC}$ is found to respond linearly to changes in the dipole tilt angle. The results for the extreme tilt angles are considered less reliable due to data coverage and are shown in transparent color. Due to a bias in $E_{sw}$ between the different dipole tilt intervals, we have normalized $\Delta\Phi_{RC}$ to represent the situation when $E_{sw} = 1.5$ mV/m, shown here in green. The results from Figure 3 are shown for reference in red.

In Figure 4 we show the underlying statistics that was used to make Figure 3. The three columns refer to the same tilt angle intervals as in Figure 3. The top row in Figure 4 shows the magnitude of the binned average $\vec{E}$ in color. $\vec{E}$ is also shown as a vector in white color in the center of each grid cell. Below each panel is also indicated the total number of LOS vectors going into the analysis. The second row in Figure 4 shows the magnitude of the estimated $\vec{E}_{SECS}$ in color and the corresponding vector as a red pin originating at the grid cell center. This is the same location as the binned average $\vec{E}$, which is also shown in these panels for reference. The SECS node locations are also shown as orange dots, located at the same latitude but half way between the locations where we evaluate $\vec{E}_{SECS}$. In the bottom panel we show the data coverage as the number of unique hours of observation in each grid cell. The color scale used is such that the most blue bins have observations from less than 50 unique hours, and therefore, its corresponding binned average $\vec{E}$ is down-weighted in the inversion as described in Paper I. Hence, the influence of grid cells showing very large $\vec{E}$ toward lower latitudes in the upper row is reduced, as is seen in the middle row in Figure 4.

The upper and middle rows of Figure 4 show that $\vec{E}_{SECS}$ is less than the binned average $\vec{E}$ in most grid cells. This is likely one of the reasons why our potentials are slightly lower (typically ∼5 kV) than the ones derived by Thomas and Shepherd (2018); see Figure 4 in Paper I. Also, the effect of our weakly imposed boundary condition is seen as $\vec{E}_{SECS}$ approaches 0 toward 60° MLAT in all MLT sectors.

## 4. Discussion

The most comprehensive investigations of seasonal variations in sunward convection velocities associated with lobe reconnection are the studies by Wilder et al. (2010) and Koustov et al. (2017). Both studies focused on the sunward convection velocity inside the polar cap during purely northward IMF, in response to the $E_{sw}$. Both studies found that on average, the summer hemisphere had stronger sunward convection velocities than winter. However, Koustov et al. (2017) found a more prominent increase in sunward convection velocities with increasing $E_{sw}$ than Wilder et al. (2010). By making assumptions about the extent and the velocity distribution across the sunward convection channel, one can approximate a measure of the associated potentials due to their observed sunward convection velocities within the $E_{sw}$ interval used in this study. Based on our results in Figure 3, a typical width of the sunward convection channel is ∼1,000 km. Using this width, and assuming the sunward convection velocities from Figures 5–7 in Wilder et al. (2010) stay constant across this distance, their results correspond to potentials of 13, 18, and 18 kV during their monthly binning that is similar to our separation into winter, equinox, and summer, respectively. We can do the same comparison using the sunward velocities in the dayside polar cap during northward IMF based on the data from Table 1 in Koustov et al. (2017). Their velocities then correspond to potentials of 4, 7, and 14 kV during winter, equinox, and summer, respectively. Our values from the bottom row in Figure 3 (9, 12, and 15 kV) are somewhere in between.

From the results presented in Figure 3 it is difficult to tell if the increase in the reverse convection potential difference $\Delta\Phi_{RC}$ is linear, as we only show three rather wide dipole tilt intervals. We have made an attempt to more accurately describe the variation in $\Delta\Phi_{RC}$ versus dipole tilt angle by performing the same analysis on smaller tilt angle intervals. A big challenge when reducing the tilt angle interval is to get enough data points to make a good determination of the binned average $\vec{E}$ vectors. In Figure 4 one can see that in the regions having observations from >50 unique hours, the binned averages of $\vec{E}$ show systematic variations between neighboring grid cells, which indicates that the binned averages represent the typical vector for the given conditions. We were able to produce similarly robust binned average $\bar{E}$ in 10° wide dipole tilt intervals between −25° and 25° (five bins). For the edge intervals (tilt > |25°|) the fit was too poor to produce a reliable result. We also increased the $E_{sw}$ interval by 0.5 mV/m in both directions to improve data coverage. The IMF clock angle interval and stability were kept the same. The results are summarized in Figure 5 as blue dots with a corresponding fitted line. The less reliable results for the dipole tilt angles >|25°| are also shown but indicated with transparent blue color. A linear increase in $\Delta\Phi_{RC}$ is seen, except for the [25°, 35°] interval, which has the poorest data coverage ($3 \cdot 10^5$ LOS vectors). Although this value has a larger uncertainty than the the other data points in Figure 5, we cannot rule out the possibility that there could be some saturation of the dipole tilt influence on $\Delta\Phi_{RC}$ for very large dipole tilt angles.

Using backscattered HF radio signals to measure the global ionospheric convection (as we do with Super-DARN) has intrinsic caveats and limitations. A successful observation depends both on the HF radio propagation conditions as well as the existence of ionospheric decameter irregularities in the electron density. Some of these issues can be seen in the statistics presented in this study and can influence the results. One example is that SuperDARN receives less backscatter echoes when the ionosphere is sunlit, reducing the amount of data for increasing dipole tilt angle. This is usually interpreted as a consequence of the sunlit plasma having weaker decameter-scale irregularities than what is needed to produce a detectable backscattered echo (Ghezelbash et al., 2014; Ruohoniemi & Greenwald, 1997), but the HF propagation conditions are also affected (Milan et al., 1997) leading to a different occurrence distribution of backscatter echoes as seen in the bottom row in Figure 4. However, due to the good radar coverage in the Northern Hemisphere, we are still able to reproduce ionospheric convection patterns during summer conditions, so this effect has likely a minor influence on the results. In addition to sunlight, we have experienced that geomagnetic activity has a similar influence on production of irregularities in the dayside polar cap. In our analysis, the data obtained in the dayside polar cap during sunlit conditions are obtained from times that are slightly more disturbed compared to the analysis when the same region is less illuminated. For the analysis in Figure 5, the mean $E_{sw}$ in the dayside polar cap increased from 1.2 to 1.7 mV/m from the lowest to highest tilt angle interval. This effect is more pronounced compared to the analysis presented in Figure 3, and it likely affects the slope of the blue line in Figure 5. Based on the results from Sundberg et al. (2009), an increase in $E_{sw}$ of 0.5 mV/m is associated with an increase of ∼4 kV in the reverse convection potential difference. Incorporating that influence by scaling the results to $E_{sw}$ = 1.5 mV/m still leads to a linear trend, as seen by the green dots and its fitted line in Figure 5. We also show the results from Figure 3 for reference as red dots and a fitted line. We can see that the Sundberg et al. (2009) correction to the blue line places it close to the results from Figure 3 (red line) that did not have a significant bias in $E_{sw}$.

Chisham et al. (2004) presented a detailed examination of the ionospheric convection in the dayside polar cap from both hemispheres during an event when IMF was stable northward. Similar to this study, they assumed a strong coupling between the ionosphere and the lobe reconnection site and related the ionospheric convection to the lobe reconnection rate. From a low Earth orbiting satellite with a favorable orbital configuration, they estimated the northern and southern reverse convection potential, interpreted as the lobe reconnection rate in the respective hemisphere, to be 13.5 and 19.7 kV, respectively. During these observations, $E_{sw}$ = 1.8 mV/m and dipole tilt = −9°. This hemispheric difference represents an increase of 46 % from the local winter reconnection rate. From Figure 5, the reverse convection potential difference $\Delta\Phi_{RC}$ is 34 % larger for dipole tilt = +10° compared to dipole tilt = −10° when using the corrected values (green dots), and 50 % larger if considering the values with a slight bias in $E_{sw}$ (1.4 mV/m vs. 1.6 mV/m, blue dots). Although our statistical averages are similar to the event based values of the lobe reconnection rate as presented by Chisham et al. (2004), deviations are expected when looking at single events, especially if the IMF direction is varying.

As illustrated in Figures 1b and 1c, we argue that when the magnetosphere and ionosphere are in equilibrium, the average ionospheric convection is largely independent of the conductivity. By first principles, the two-cell flux transport is controlled by the dayside/nightside reconnection rates and should be equal in both hemispheres. Hence, the two-cell convection pattern in the two hemispheres should on average be similar in terms of magnetic flux transport. However, from observations, this interpretation can in some cases be challenging to justify. Chisham et al. (2009) presented average maps of the vorticity of the ionospheric convection deduced from SuperDARN measurements. They found a seasonal dependence of the vorticity, where stronger vorticity was found during summer compared to winter. This seems to contradict that the two hemispheres are highly coupled and that the convection is to a large degree similar in the two hemispheres. However, as discussed earlier in this section, there are inherent limitations of the SuperDARN technique when comparing average convection from different seasons. From Figures 6 and 8 in Chisham et al. (2009) it is evident that the entire oval region is shifted to lower latitudes during the summer statistics, indicating that the underlying data are sampled during higher levels of activity making the summer/winter comparison challenging. Also Pettigrew et al. (2010) found, using SuperDARN, a tendency of larger cross polar cap potential in the summer hemisphere, also during southward IMF when no lobe reconnection is expected to occur. Despite sorting the observations by $B_T$ (IMF magnitude in GSM YZ plane), the results from Pettigrew et al. (2010) are likely affected by the geomagnetic activity bias mention above, highlighting the difficulty to asses the conductivity influence on the ionospheric convection also in that study. In a recently developed model of the ionospheric convection using SuperDARN (Thomas & Shepherd, 2018), the Northern Hemisphere cross polar cap potential difference was found to vary little with season during southward IMF. In that study, the underlying data were sorted by $E_{sw}$ in 0.5 mV/m intervals, likely further reducing the influence of the activity bias in SuperDARN compared to the study by Pettigrew et al. (2010). Furthermore, in a study of the nightside convection velocities focusing on the return flow (Reistad et al., 2018), no significant seasonal difference below 70° MLAT and away from the nightside convection throat can be seen in the return flow when $AL > -150$ nT. We therefore conclude that the observed variations in ionospheric convection with dipole tilt is mainly reflecting the changing levels of external driving, namely, the lobe reconnection rate when focusing on the dayside polar cap. Note that an underlying assumption for this strong coupling is that there are no parallel electric field in this region.

In Figure 2 we presented numbers reflecting the strength of the NBZ current system during northward IMF from the AMPS model, for three different values of the dipole tilt angle corresponding to the intervals used in Figure 3. The AMPS model indicates that the average NBZ currents are approximately four times stronger in summer than in winter (dipole tilt angle 21° vs. −15°). Since our results of the reverse convection above 80° MLAT on the dayside is considered independent of the ionospheric conductivity, at least to the first order, we can make a quantitative estimate of how much this increase in field-aligned current is attributed to the increase in solar-induced ionospheric conductivity. To do so, we need to assume that the conductivity in this region (dayside polar cap) is uniform, simplifying the relation between the Birkeland current, the Pedersen conductance, and $\Delta n_e$, as expressed in equation 16 in Paper I. Based on the minimum values of $\Delta n_e$ from Figure 3 corresponding to the "+" location in Figure 2, we use equation 16 in Paper I to compute the Pedersen conductance. We obtain values of $\Sigma_P$ of 1.8, 3.1, and 4.6 S for winter, equinox, and summer, respectively, indicating an increase in $\Sigma_P$ of a factor of 2.6 when the field-aligned current increases by a factor of 3.8. Hence, the increase in NBZ currents with increasing dipole tilt angle is mostly an effect of the increased ionospheric conductivity, highlighting the shortcoming of using currents alone to quantify the magnetospheric source process for this purpose. Comparing to $\Sigma_P$ estimated by the empirical model by Moen and Brekke (1993) using the $F$10.7 solar flux and the solar zenith angle, we get 8.1 S during the summer conditions, 3.3 S during equinox, and 0 during winter as the model go to zero when the solar zenith angle reach 90°.

Recently, Laundal, Reistad, et al. (2018) reported that when IMF was southward, the combination of IMF $B_x$ and dipole tilt angle could modify the dayside reconnection rate. For northward IMF, an IMF $B_x$ influence has been speculated, but results are ambiguous (Østgaard et al., 2003; Yakymenko et al., 2018). We have also looked into the possible influence of the sign of the IMF $B_x$ component on the reverse convection potential difference, $\Delta\Phi_{RC}$. When separating the analysis above into positive and negative IMF $B_x$, $\Delta\Phi_{RC}$ was found to be very similar to those presented in Figure 3. When using observations only during negative IMF $B_x$, we obtained values for $\Delta\Phi_{RC}$ of 10, 13, and 15 kV, while during positive IMF $B_x$ the corresponding $\Delta\Phi_{RC}$ was found to be 9, 11, and 14 kV. Hence, we conclude that the IMF $B_x$ influence on $\Delta\Phi_{RC}$ and therefore lobe reconnection rate is small compared to the dipole tilt influence. These results are consistent with the findings of Yakymenko et al. (2018) that reported a similar response of the sunward convection speed in the polar cap to $E_{sw}$ during both positive and negative IMF $B_x$ conditions. Despite being a small effect, the difference in $\Delta\Phi_{RC}$ between positive and negative IMF $B_x$ conditions is in the expected direction (Østgaard et al., 2003) in each of the three dipole tilt angle intervals.

This study aims to quantify the influence of the dipole tilt angle on the lobe reconnection rate. Other reconnection geometries than the IMF-lobe reconnection scenario (Figure 1a) has been suggested to play a role during northward IMF (Watanabe et al., 2005; Watanabe & Sofko, 2009a, 2009b), possibly complicating the interpretation of $\Delta\Phi_{RC}$ as the lobe reconnection rate. In particular, the so-called "interchange cycle" (Watanabe & Sofko, 2009a) is expected to lead to reverse convection on closed field lines in the hemisphere opposite to where IMF-lobe reconnection takes place. It is therefore possible that our observations during winter are influenced by the flux transport being part of the interchange cycle due to lobe reconnection in the opposite summer hemisphere, since we are not explicitly distinguishing between open and closed field lines in our analysis. However, our analysis indicates that the source region of the reverse convection also in the winter is above 80° MLAT, usually interpreted as open field lines. Then, the inferred reverse convection potential difference $\Delta\Phi_{RC}$ can be interpreted as the lobe reconnection rate in the hemisphere we do the analysis (assuming "reverse Dungey" type reconnection—Watanabe & Sofko, 2009b—is negligible). If this region is in fact threaded by closed field lines, and lobe reconnection (or IMF-closed reconnection; Watanabe & Sofko, 2009b) is exclusively a summer phenomenon, as suggested by Watanabe and Sofko (2009b), our winter results must be interpreted as the fraction of the IMF-lobe reconnection from the opposite hemisphere that participate in the interchange (and/or reverse Dungey; Watanabe & Sofko, 2009b) cycle, while the summer results is the IMF-lobe (and possibly reverse Dungey) reconnection rate in the summer hemisphere.

Regardless of the importance of the additional high-latitude reconnection scenarios during northward IMF, the dipole tilt influence on the lobe reconnection rate is highly significant. In light of the above discussion, interpreting the observed reverse convection potential difference, $\Delta\Phi_{RC}$ as the lobe reconnection rate will place a lower limit of the influence of dipole tilt on the lobe reconnection rate in the local hemisphere. Any contribution from the interchange and/or reverse Dungey cycle reconnection (Watanabe & Sofko, 2009b) will lead to a stronger influence of dipole tilt on lobe reconnection rate in the Northern Hemisphere compared to the trend seen in Figure 5. Hence, the importance of the reconnection geometries during northward IMF suggested by Watanabe et al. (2005) and (Watanabe & Sofko, 2009a, 2009b) needs further investigations to more accurately address the lobe reconnection rate in each hemisphere separately.

Our results suggest that the two main ingredients in a lobe reconnection coupling function during northward IMF are most likely $E_{sw}$ and the dipole tilt angle. The implication of this significant dipole tilt dependence is that the electrodynamics in the northern and southern polar cap regions can be severely different. These are important global differences that at the moment are not well quantified and understood. Deriving a first-order global lobe reconnection coupling function will be a first step of quantifying this hemispheric difference due to the hemispheric differences in the lobe reconnection rate. As demonstrated here, the SECS technique is well suited for such further studies.

## 5. Conclusions

In this paper we have demonstrated the use of the SECS representation of the average ionospheric convection electric field, as described in Paper I. The new ability to separate and quantify the sources of the ionospheric convection is shown to be highly applicable for a quantitative description of the reverse convection potential during northward IMF. The findings in this paper can be summarized as follows:

- The dipole tilt angle is a secondary important controlling parameter of the lobe reconnection rate during northward IMF, after $v_{sw}B_z$.
- During northward IMF, the Northern Hemisphere typically has a reverse convection potential difference during summer that is two times that during winter, suggesting a strong hemispheric asymmetry. This influence can be considered as a lower bound of the dipole tilt influence on lobe reconnection rate in each hemisphere during northward IMF.
- The reverse convection potential difference depends linearly on the dipole tilt angle in the range [−25°, 25°].
- The SECS representation of the ionospheric convection allows for a convenient separation of the inferred magnetospheric sources of the convection, making it highly suited for further studies of the controlling parameters of lobe reconnection.

**Acknowledgments**
SuperDARN (Super Dual Auroral Radar Network) is an international collaboration involving more than 30 low-power HF radars that are operated and funded by universities and research organizations in Australia, Canada, China, France, Italy, Japan, Norway, South Africa, the United Kingdom, and the United States. The SuperDARN data were obtained directly from Evan Thomas, but raw files can be accessed via the SuperDARN data mirrors hosted by the British Antarctic Survey (https://www.bas.ac.uk/project/superdarn/#data) and University of Saskatchewan (https://superdarn.ca). We acknowledge the use of NASA/GSFC's Space Physics Data fFacility (http://omniweb.gsfc.nasa.gov) for OMNI data. Financial support has also been provided to the authors by the Research Council of Norway under the contract 223252.